\documentstyle[epsfig,11pt]{article}     
\begin{document}

\noindent

\begin{center}

\large 

{\bf ``Bernoulli" Levitation}

Chris Waltham, Sarah Bendall and Andrzej Kotlicki 

{\it Department of Physics and Astronomy, 
University of British Columbia, Vancouver B.C., Canada V6T 1Z1}

\normalsize

Email: waltham@physics.ubc.ca

Revised for submission to Am. J. Phys. (2002/06/10)

\end{center}

\section*{Abstract}

``Bernoulli" levitation is the basis of many popular counter-intuitive
physics demonstrations. However, few of these lend themselves to a
quantitative description without recourse to computational fluid dynamics.
Levitation of a flat plate is the exception, and we present here a
straightforward analysis which illustrates several principles of fluid
mechanics in a pedagogically useful way.

\section{Introduction}

The reduction of pressure in a moving stream of fluid is commonly called
the Bernoulli Effect. It can be simply and dramatically demonstrated by
blowing air into the narrow end of a small funnel. A ping-pong ball placed
in the funnel cannot be blown out; the harder one blows, the more it
sticks in place.  A quantitative description of this
demonstration is not easy however, due to the complicated geometry and
airflow.

To avoid these complexities, we have produced produced a version of 
this
demonstration which is even more dramatic than the ping-pong ball, and yet
lends itself to a quantitative description.  Air is blown 
vertically downwards through a 
hose which exits in a flat horizontal sheet. Another flat sheet brought 
up to the orifice will be held in place despite the fact that the air 
is pushing downwards. The acceleration of the air in the gap causes a 
drop in pressure which more than compensates for the high pressure in the 
hose. Flowing liquids will also produce the same effect, as one can easily 
show using a flat sheet placed against a water jet in a swimming 
pool or hot tub.

Our design (figs. 
1 and 2) was fabricated from 3/8" acrylic sheet, a material chosen for 
its
transparency and the smoothness of its surface. The design is not
original; it probably exists in thousands of versions. The authors
were inspired by an exhibit at the Pearson Air Museum in Vancouver,
Washington\cite{Pearson}. What we show here is that the system is amenable
to a fairly
straightforward analysis which illustrates various principles of fluid
motion. 

The source of air is an outlet of a  common half-horsepower shop vac 
(which has never been used for cleaning). The nozzle (inner radius 
$r_{hose}$ = 
14~mm) 
is 
inserted into a hole in a horizontal sheet of acrylic, blowing downwards.
The edge of the hole is rounded with an approximately 1~mm radius of 
curvature.
A flat disk of 3/8" acrylic (radius 150~mm)  brought up to the hole at 
first experiences a 
strong downward force, but on being pushed closer it is suddenly grabbed 
by 
the air flow and held in place about 1~mm below the fixed sheet of 
acrylic. This arrangement can levitate approximately 2~kg. In our 
apparatus, the pressure can be measured in the hose and at three different 
radii above the suspended plate. The most important two pressures to 
measure 
are the one  in the hose and that just outside the hose radius, where the 
air velocity 
is the highest and the pressure the lowest. 
It is the thin disk 
of high 
velocity air just outside the hose which is  responsible for the 
levitation.
The pressures are displayed 
by two large analog gauges, chosen for visual effect. 
The hose pressure is controlled using a Variac power supply for the shop 
vac.
   
\section{The Calculation}

The air flows relatively slowly down the hose and then is forced into the 
small gap above the plate. Mass conservation dictates that the air speeds 
up considerably here, and Bernoulli's equation gives the associated 
pressure drop.

\begin{equation}
P_1 + \frac{1}{2}\rho_1 v_1^2 + \rho_1 g h_1 = 
P_2 + \frac{1}{2}\rho_2 v_2^2 + \rho_2 g h_2 
\end{equation}

Here the subscripts 1 and 2 refer to the hose and the entrance to the 
channel respectively. The
pressure is $P$, density $\rho \approx 1.24$kg/m$^2$ (in our cool 
laboratory), velocity $v$, height
$h$ and the gravitational acceleration is $g$. The velocities are 
calculated from conservation of mass. As we will see, a typical 
value of
$P$ is several kPa and $v > 10$~m/s, and so the last terms involving 
height
variations of a few cm can be safely ignored. As will be demonstrated 
below, the flow is turbulent and so has no thick boundary layer. 
Hence the air velocity is approximately uniform across the width of the 
channel. The velocity leaving the hose approaches 200 m/s. This 
is rather larger than  Mach 0.3 ($\sim 100$~m/s), 
which is usually considered to be the velocity above which compressibility 
has to be taken into account. We shall show the small but significant 
effects of 
compressibility.

For the flow between the two plates, consider a wedge of air between 
as shown in Fig. 2a.
Its radial extent is from $r$ to $r+dr$, azimuthal extent $\delta$, and a 
constant thickness 
 $x$.
Air flows through the wedge at rate of $dq$~kg/s from left to right, 
entering with velocity $v$ and pressure $p$ and leaving with velocity 
$v+dv$ 
and pressure $p+dp$. We can solve the momentum equation for the wedge
by considering the two forces, one from the pressure difference and one 
from wall friction, which cause the acceleration of the air passing 
through.
The force due to air pressure acts on the four sides of the wedge. On the 
left hand edge the total force is:

\begin{equation}
F_{l} = prx\cdot 2\sin\frac{\delta}{2}
\end{equation}

Here the positive direction is to the right and the factor 
$2\sin\frac{\delta}{2}$ 
arises from the small effect of 
curvature. 
Similarly, the force on the right hand edge is:

\begin{equation}
F_{r} = -(p+dp)(r+dr) x\cdot (2\sin\frac{\delta}{2})
\end{equation}

The radial contribution of the two straight sides cancels out terms in 
$pdr$:

\begin{equation}
F_{s} = (p+\frac{dp}{2})(dr)x\cdot (2\sin\frac{\delta}{2})
\end{equation}

Hence the total contribution from the pressure only has terms in
$rdp$:

\begin{equation}
F_{p} = F_{l}+F_{r}+F_{s} = -rxdp\cdot (2\sin\frac{\delta}{2}) 
\end{equation}

The force due to wall friction is given in terms of
the friction factor $f$, which can be applied to flow
between flat plates as follows\cite{White}:

\begin{equation}
F_{f} = -xrdr\frac{f}{4x} \rho v^2\cdot (2\sin\frac{\delta}{2})
\end{equation} 

The friction factor can be either be read off a Moody Chart if 
the Reynolds number is fixed and known, or the factor can be 
calculated 
from the unlikely empirical expression for turbulent flow\cite{White}
which is plotted in fig.~4:

\begin{equation}
\frac{1}{\sqrt{f}} = 2.0 log_{10}(0.64 Re_{D_h} \sqrt{f}) - 0.8
\end{equation}  

The characteristic Reynolds Number for channel flow,
$Re_{D_h}$,
uses the characteristic length $D_h$ defined to be 4 times the
 area of the channel divided by the wetted perimeter, i.e.
twice the gap distance.

\begin{equation}
 Re_{D_h} = \frac{\rho v D_h}{\mu}, \ \ \ \ D_h \approx 2x
\end{equation}

The fluid density, viscosity and velocity are given by $\rho$, $\mu$ and $v$.
As the variation of the friction factor $f$ with $Re$ is logarithmic, we
will use this approximation throughout. The Reynolds number  
varies by an order of magnitude (2,000-20,000) in this system, but $f$ 
only
varies between 0.02 and 0.05. For the most important region of flow, just 
outside the hose, $f\approx 0.023$.

The rate of change of momentum of air passing through the wedge can be 
found 
using the mass flow in the wedge $dq$ and the change in velocity $dv$.

\begin{equation}
\frac{d(m\vec{v})}{dt} = dq d\vec{v} = q dv\cdot (2\sin\frac{\delta}{2})/2\pi 
=F_p+F_f
\end{equation}

The total mass flow $q=2\pi r x \rho v$, and the incremental velocity change 
$dv$ is simply related to $dr$ because
of mass conservation and the cylindrical geometry:

\begin{equation}
\frac{dv}{dr} = -\frac{v}{r}
\end{equation}

Equating force and the rate of momentum change produces a very simple 
result:

\begin{equation}
dp = \rho v^2 dr \left(\frac{1}{r} - \frac{f}{4x}\right)
\end{equation}

If one allows the density to vary with pressure, the result 
is:

\begin{equation}
dp = \frac{\rho_a v^2 dr \left(\frac{1}{r} - \frac{f}{4x}\right)
\left(\frac{p}{p_a}\right)^{1/\gamma}}
{\left(1 - \frac{\rho_a v^2}{\gamma 
p_a}\left(\frac{p}{p_a}\right)^{1/\gamma}\right)}
\end{equation}

The subscript $a$ here refers to ambient values, and $\gamma$ can be set 
to 1 for isothermal density variations and 1.4 for adiabatic variations.
In practice, there is very little discernible difference between these two 
cases.

The pressure was integrated numerically using radial steps of 0.1~mm
starting at the hose radius. At each step the friction factor was
evaluated from the local Reynolds number. The calculation was done in
Excel, using the ``solver"  utility to find the right value of the 
total flow rate $q$ for a
measured hose pressure which yielded the ambient pressure as the air left
the channel. The gap size was then varied and the total force on the 
plate was found. The minimization procedure was often not 
straightforward. In 
some cases quite different solutions 
could 
be found using different starting guesses;
the maximum pressure changes sometimes differed by 
as much as 10\%, and the force on the plate by several Newtons. 
This may be due to the fact that 
the force in the plate is, in reality, quite a small difference between 
two 
much larger forces on either side.

\section{Levitation Measurements}

The acrylic plate we used for levitation had a mass of 785~g (769~N).
Sometimes an additional 1~kg mass was added, which was the limit for
the shop vac used. 
Measurements of the gap were taken at various hose pressures (3-10~kPa). 
The plate was stable enough to use a simple Feeler 
gauge. Pressure measurements were  
made at a radius of 20~mm, close to the minimum pressure region.
The ambient temperature was 5C and the pressure 100~kPa.


A problem encountered in the measurements is a rocking motion at a few Hz 
caused by 
the fact that the plate is only supported by a thin ring of air 
just outside the hose radius. This can be cured after onset by lightly 
touching the plate with a finger. It could presumably be reduced by using 
a 
smaller plate.

A interesting effect occurs when the additional mass of 1~kg is hung 
from the plate. With this weight, the plate tends to float crookedly, 
with a gap of 1.55~mm on one side and 1.95~mm on the other. This is 
presumably a kind of Euler instability which bears further investigation 
in the future.

\section{Comparison of Data and Calculations}

Figs. 5-7 show calculations made with a representative hose pressure of
8~kPa. At this pressure the gap was measured to be $x = 1.00\pm 0.05$~mm.
Fig. 5 shows how the pressure distribution depends on 
different assumptions. 
The calculation which best reproduces the total 
upward force on the plate (769~N) is the isothermal approximation with the 
nominal friction factor and a 1.1~mm gap. The adiabatic approximation 
yields an almost 
identical result, and the measurements cannot distinguish between the two.
However, an incompressible flow yields a slightly broader, 
shallower 
pressure profile and requires a larger gap (1.3~mm) to obtain the right force. 
This is not so consistent with the data as the isothermal/adiabatic case. Also 
shown in the plot 
is the frictionless case for a 1.1~mm gap, which gives almost double the 
measured pressure drop and four times the upward force.


Fig. 6 shows the air velocity as the air exits the hose.
The gap is 1.1~mm and the flow isothermal. The velocity peaks 
at almost 200~m/s shortly after entering the gap. At the same place, the 
pressure 
reaches a minimum of -10~kPa, as seen in fig. 5. 
The Reynolds number distribution with 
radius is shown in  fig. 7; the peak value is nearly 30,000. 

The gap measurements are presented in fig.~8, with and without the extra 
1~kg mass suspended under the floating plate. The agreement with the 
calculation is good, although the small non-linearity in the theoretical 
curves is most
likely a result of the calculational problems noted above, and not a
real physical effect.

There is a second, larger gap size which produces a  small upward 
aerodynamic force sufficient to lift
the plate. However this position is unstable as the
upward force rises as the gap size is reduced, and so it cannot be used 
for levitation.  Experimentally it occurs
at around $x \sim 7$~mm.
One can easily estimate this  gap size 
by 
requiring that the 
cross-sectional area 
does not change between the hose and the channel between the plates:

\begin{equation}
\pi r^2_{hose} = 2\pi r_{hose}x \ ; \ \ \ i.e. \ \ 
x=r_{hose}/2=7\mbox{ mm}
\end{equation}

In this case there can be no drop in pressure above the plate, and the 
force
will be zero. A small decrease in this gap will support the plate in unstable
equilibrium.

\section{Conclusions}

We have produced a simple and spectacular demonstration
of Bernoulli
levitation which can be used in front of a large audience.
We have shown that 
it is possible, simply by using momentum conservation and 
common friction factors,
 to account quantitatively for all the main features of the demonstration.

\section*{Acknowledgements}

This demonstration was designed by one of the authors (SB) while taking the
Physics 420 ``Physics Demonstrations" course at the University of British
Columbia (UBC) Department of Physics and Astronomy.  The apparatus was made by
Philip Akers of the departmental machine shop. The authors thank the UBC
Teaching and Learning Enhancement Fund for supporting this course. Thanks also
to Professors Emeritii Boye Ahlborn and Douglas Beder for illuminating
discussions on the infinite subtleties of fluid flow, to Robert Waltham for
showing us that submerged water jets in a swimming pool can also levitate
large objects, and to Susanna and Christine Waltham for help with the
photography.

\pagebreak

\section{Figures}

Figure 1a: General arrangement of the demonstration, showing the acrylic
frame, gauges, levitation plate, Variac and shop vac. The surgical tubes
are used to pick off the pressure at two points: in the hose and between
the plates at a radius of 20~mm.

Figure 1b: Levitation of plate with 1~kg mass suspended below. The shop 
vac hose enters from the top in the centre. The vertical acrylic pegs 
prevent the plate from floating off to one side more than a few mm.

Figure 2: General arrangement and dimensions of Bernoulli demonstration.
The air is piped into the top from a commercial shop vac. It is fabricated
from 3/8" acrylic sheet.

Figure 3: Wedge of air between the two plates used in the calculation.

Figure 4: The friction factor $f$ as a function of Reynolds Number $Re$,
as given by eq. 7.

Figure 5: A plot of pressure versus radius for the isothermal 
approximation. Hose pressure 8~kPa, gap 1.1~mm.

Figure 6: A plot of velocity versus radius for the isothermal
approximation. Hose pressure 8~kPa, gap 1.1~mm.

Figure 7: Reynolds Number as a function of radius for the isothermal
approximation. Hose 
pressure 8~kPa, 
gap 1.1~mm.

Figure 8: Hose pressure versus gap size for two different plate weights.
The points are data, taken at an ambient temperature and pressure of 5C 
and 101~kPa respectively.

\newpage
\begin{figure}
\center{\epsfig{file=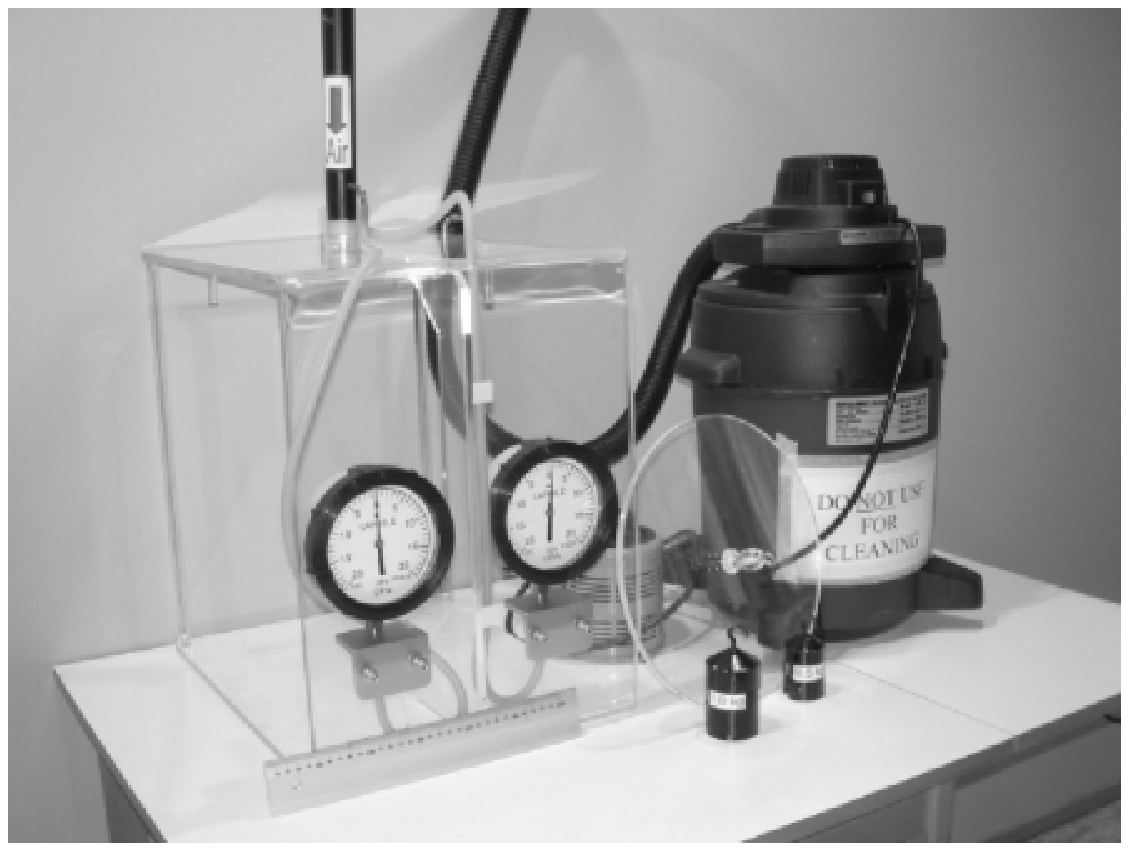,width=12cm}}
(a)
\center{\epsfig{file=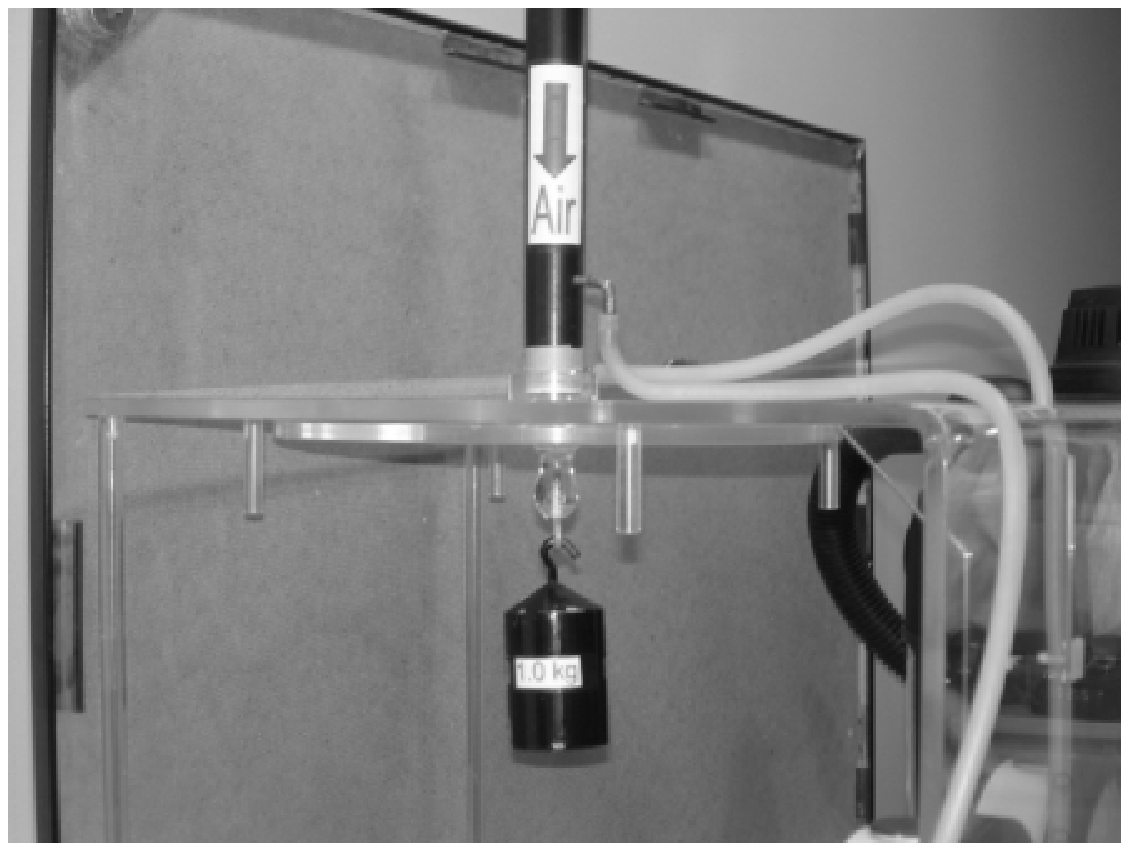,width=12cm}}
(b)
\caption{}
\end{figure}

\newpage
\begin{figure}
\center{\epsfig{file=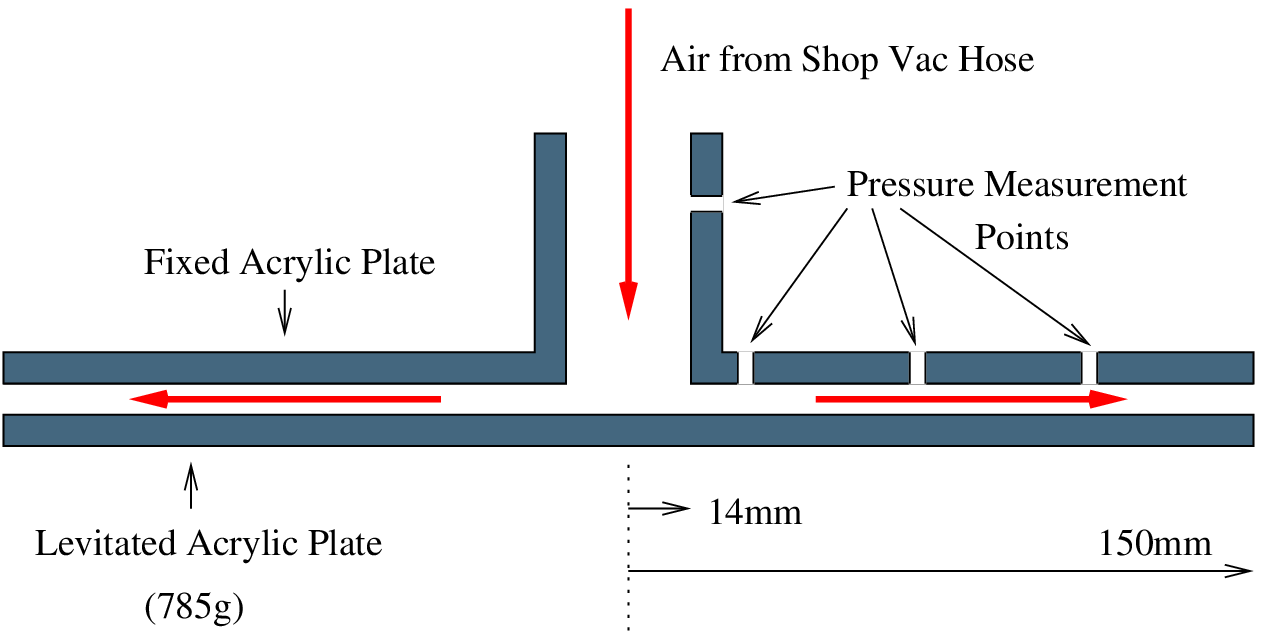,width=12cm}}
\caption{}
\end{figure}

\begin{figure}
\center{\epsfig{file=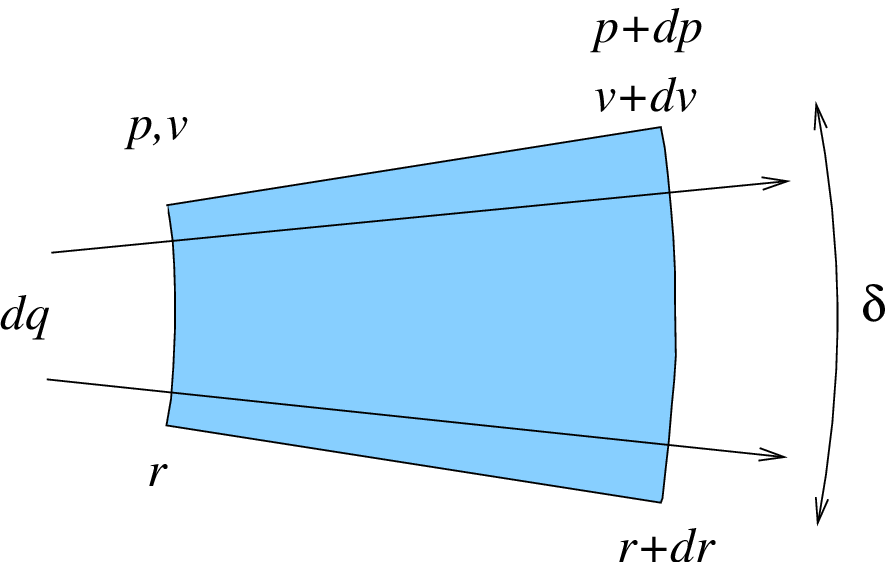,width=12cm}}
\caption{}
\end{figure}

\begin{figure}
\center{\epsfig{file=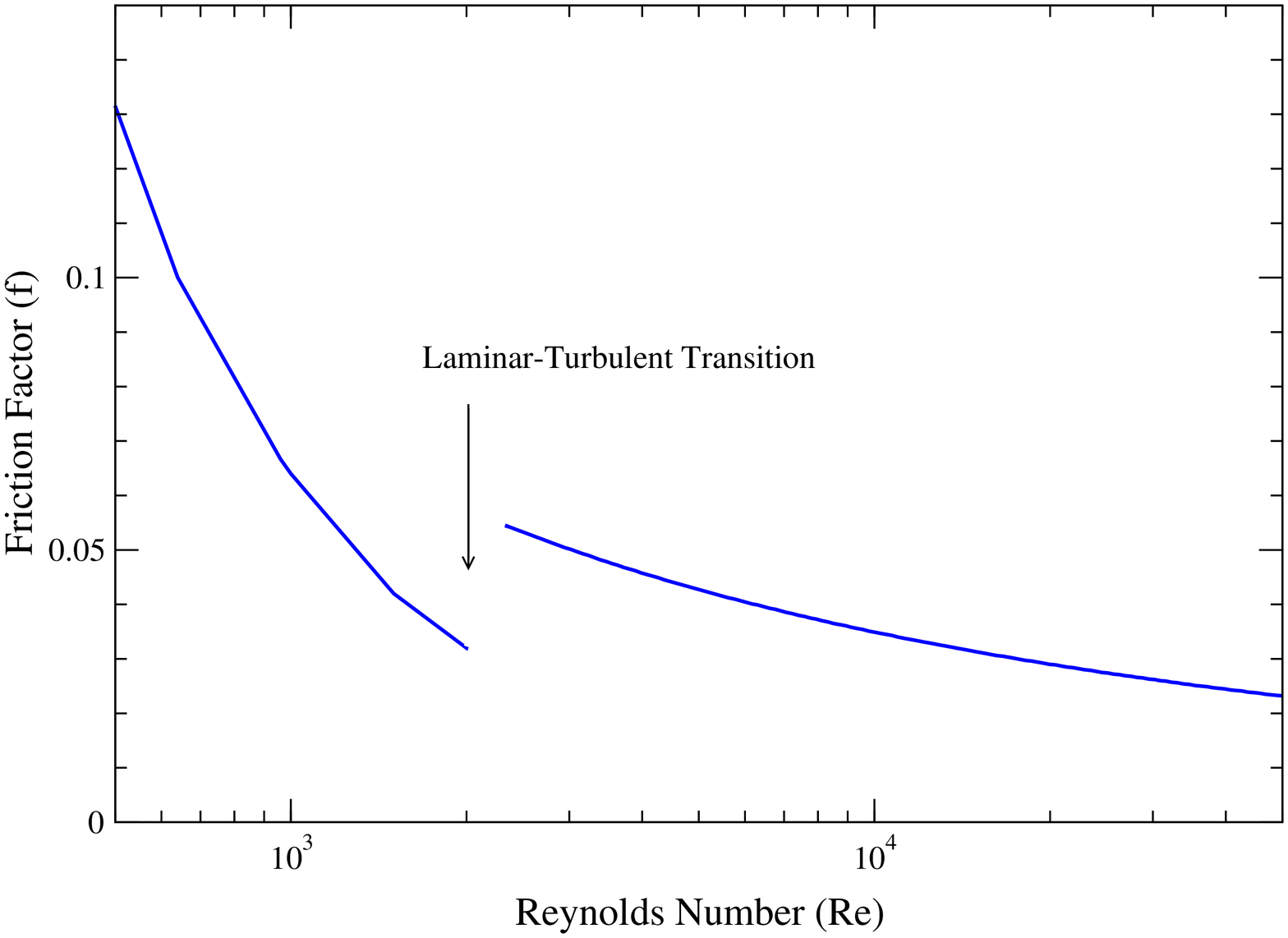,width=10cm,angle=-90}}
\caption{}
\end{figure}

\begin{figure}
\center{\epsfig{file=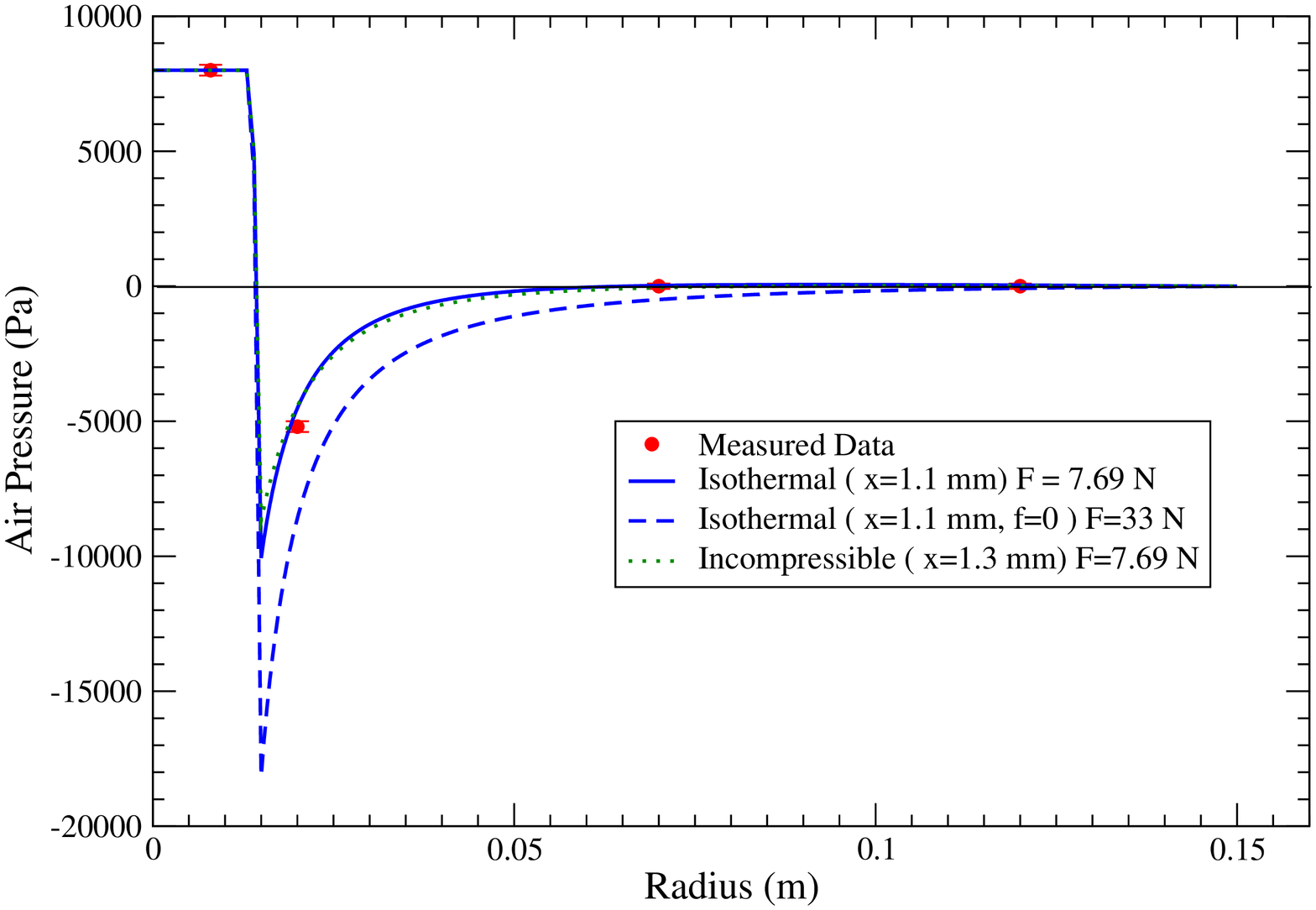,width=10cm,angle=-90}}
\caption{}
\end{figure}

\begin{figure}
\center{\epsfig{file=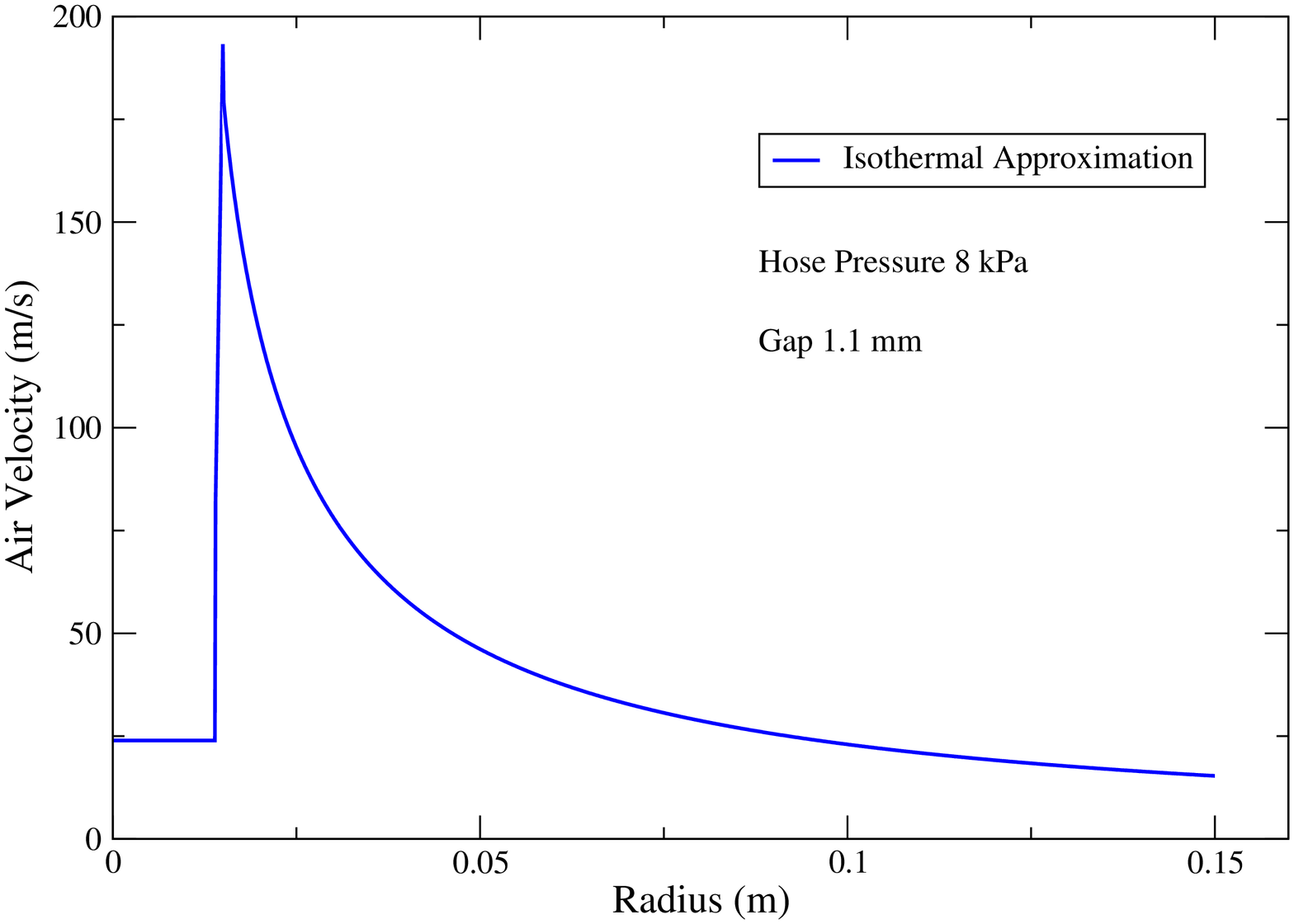,width=10cm,angle=-90}}
\caption{}
\end{figure}


\begin{figure}
\center{\epsfig{file=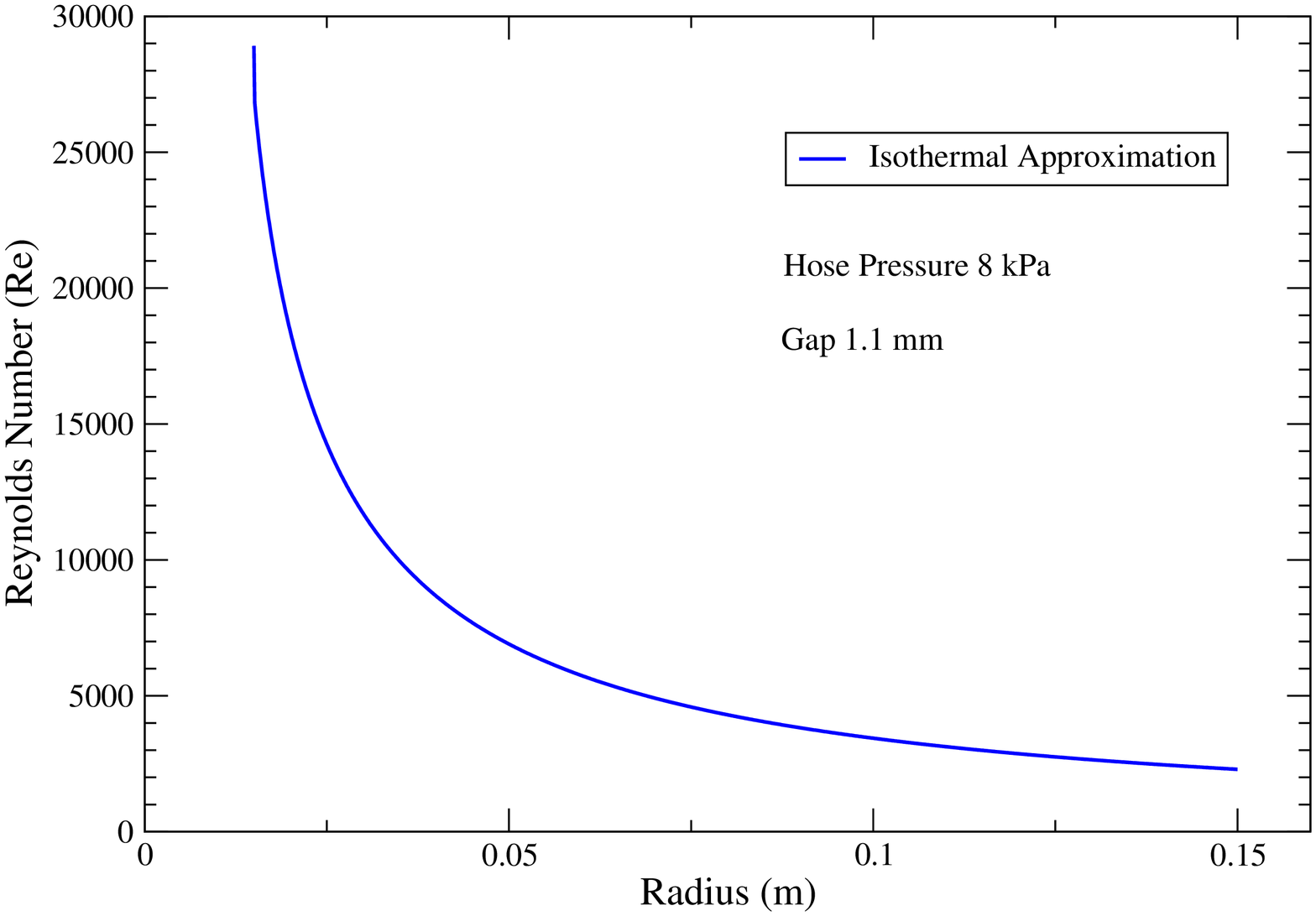,width=10cm,angle=-90}}
\caption{}
\end{figure}

\begin{figure}
\center{\epsfig{file=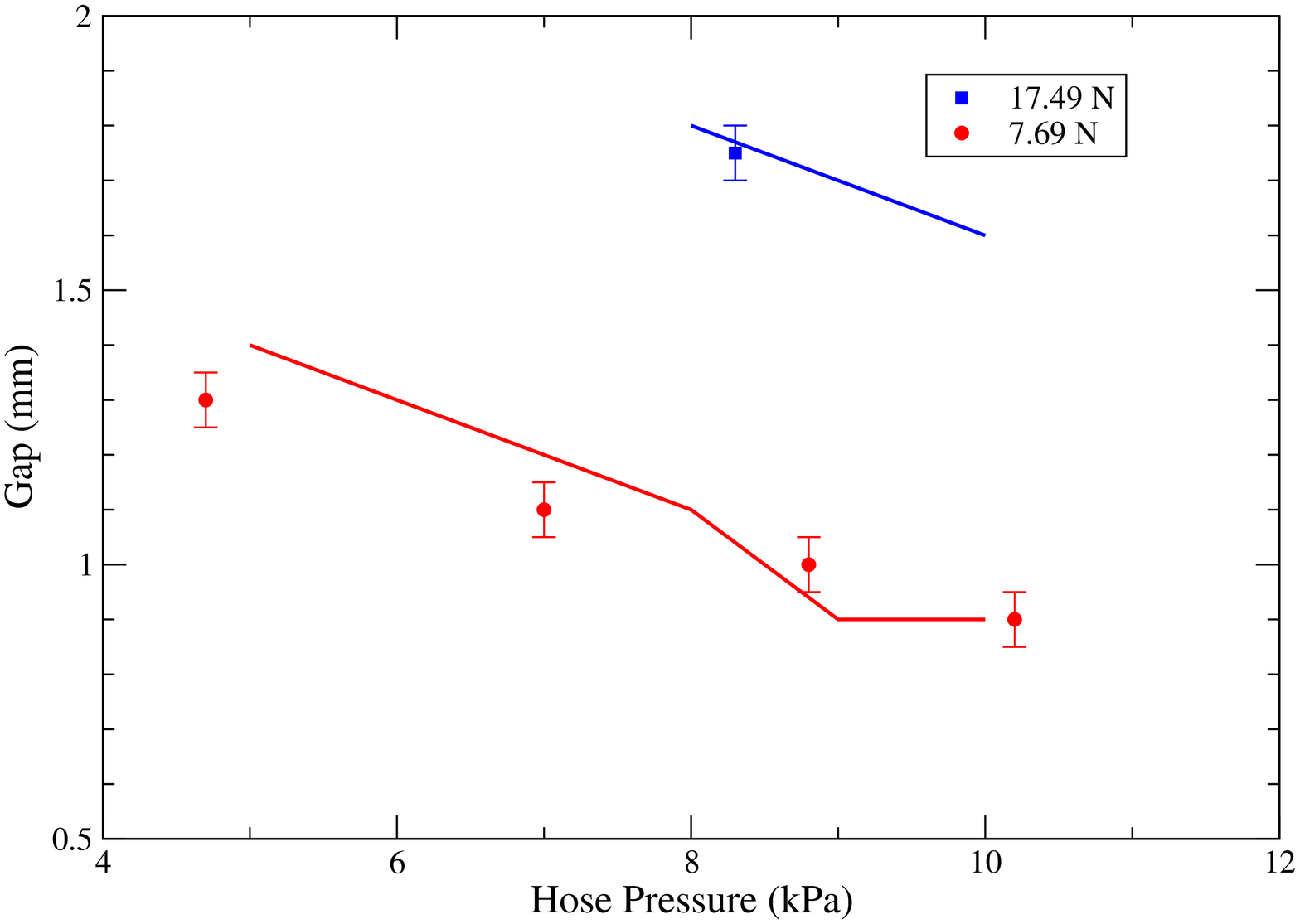,width=10cm,angle=-90}}
\caption{}
\end{figure}


\begin{thebibliography}{99}

\bibitem{Pearson} Jack Murdock Aviation Center, Pearson Field, 1105 E 5th
St, Vancouver WA; http://www.ci.vancouver.wa.us/murdock.htm

\bibitem{White} Frank M. White, {\it ``Fluid Mechanics"},  2nd 
edition, McGraw-Hill (1986), section 6, p.287ff

\end{thebibliography}
\end{document}